\documentstyle[aps,prd]{revtex}

\newcommand{\SA}[1]{\sin^{#1}\theta}
\newcommand{\CO}[1]{\cos^{#1}\theta}

\newcommand{\LL}{l(l+1)}
\newcommand{\NN}{\nonumber\\}
\begin{document}

\title{
Approximate equation relevant to axial oscillations on 
slowly rotating relativistic stars  }

\author{
Yasufumi Kojima and Masayasu Hosonuma }

\address{ 
Department of Physics, Hiroshima University,  
         Higashi-Hiroshima 739-8526, Japan  
           }

\maketitle

\begin{abstract}
Axial oscillations relevant to the r-mode instability  
are studied with slow rotation formalism in general relativity.
The approximate equation governing the oscillations 
is derived with second-order rotational corrections.
The equation contains an effective 'viscosity-like' term, which 
originates from coupling to the polar g-mode displacements.
The term plays a crucial role on the resonance point,
where the disturbance on the rotating stars satisfies a 
certain condition at the lowest order equation.
The effect is significant for newly born hot neutron stars,
which are expected to be subject to
the gravitational radiation driven instability of the r-mode.
\end{abstract}

\section{Introduction}

The surprising discovery of the r-mode instability in rotating stars
has inspired the study of the axial oscillations%
~\cite{an98,frmo98,ko98,lom98,olcsva98,beko99,koho99,liip99,%
lmo99,lofr99,yole00,ykye99,aks99,hi99,li99,bi99,le99,rls99,frlo99}.
The physical mechanism of the instability is 
the same as that in the polar modes, so-called
the radiation reaction instability found by 
Chandrasekhar~\cite{ch70}, and Friedman and Schutz~\cite{frsc78}.
The r-mode oscillations seem to be more important
since they are unstable on an inviscid rotating 
fluid even for small angular velocity. 
This instability can explain the spin-down process of newly born 
neutron stars which rotate with nearly Kepler frequency.
The gravitational waves associated with the unstable
r-mode oscillations may be promising detectable sources on the ground
based laser interferometric detectors.
It was proposed that the unstable mode might also play a key role on the
spin of the accreting white dwarfs~\cite{aks99,hi99}. However, 
Lindblom~\cite{li99} showed that the possibility is unlikely realized.
It was also proposed that the r-mode instability could 
provide the loss of angular momentum from the accretion disk
by the gravitational waves to halt the spin-up  in 
the low mass X-ray binaries~\cite{aks99,bi99}. 
The proposed conclusion crucially depends on the poorly
understood dissipation mechanism~\cite{le99}.
In this way, the r-mode oscillations enrich the astrophysical
implications.
Recent review of this subject is given by Friedman and Lockitch%
~\cite{frlo99}.

Most of the studies are however limited to idealized situation.
Some effects are added to the simplified models
to examine the validity. 
For example, Rezzolla, Lamb and Shapiro%
~\cite{rls99} suggested that the magnetic field of a neutron star
is wound up during the non-linear growth of the unstable
mode, and that the energy is not transfered to
the gravitational radiation so much.
However, their calculation is not self-consistent 
non-linear one, so that the magnetic effect on
the r-mode is not conclusive at moment.
The relativistic effects are also important for the
oscillations in neutron stars.
The relativistic factor is of order 0.2, so that the frequency
could slightly shift from the Newtonian calculation. 
More important effects of the general relativity are 
gravitational wave and frame dragging. 
Each of them leads to qualitatively a different result.
Some authors~\cite{ko98,beko99} already calculated the r-modes
in general relativity.  The frame dragging effect is regarded 
as a kind of differential  rotation ~\cite{ko98}. 
The rotational effects are however limited to the lowest order.
Mathematically, the treatment is insufficient, since
the modes are degenerate at the order. 
It is necessary to include the higher order rotational corrections. 
Such a task will be significantly complicated 
when the rotation and relativistic effects are 
simultaneously considered.

One simplification for the non-rotating case 
is realized as the decomposition of 
the spherical harmonic function.
Each oscillation mode can be specified for
each index $l, m.$
Furthermore, the polar and axial modes are
definitely determined by an appropriate combination of 
the harmonic functions.
However, the functions should be mixed in the presence of 
the rotation.
Considering the slow rotation, the coupling is weak,
so that the entangled range can be restricted.
It is not known a priori to specify the axial oscillations
on the rotating stars by a few spherical harmonic functions.
Indeed, Lockitch and Friedman~\cite{lofr99}
calculated the normal mode by the sum of 
infinite number of the spherical harmonics indices.
We will consider a different approach in this paper.
Our treatment is suitable for the initial-value problem.
We do not consider a single Fourier mode $ e^{-i \sigma  t}$  
with respect to time. 
Suppose that the initial disturbance at $ t=0 $ is produced
with a certain symmetry, 
which can be assumed to be specified by a few number of 
spherical harmonic functions.
What happens in the subsequent evolution?
If the oscillation possesses the symmetry, 
the oscillation preserves the symmetry.
Otherwise, new patterns with different spherical harmonics 
indices are induced. It is clear that 
the truncated approximation to the finite number of the
spherical harmonics 
becomes worse for large $t$ in general. 
Therefore, our method is 
constructing the pulsation equation adequate for 
small $t.$
It is difficult to address the valid domain of $t$
beforehand.
For example,
the r-mode oscillation can be well described 
by a few number of the  spherical harmonics indices
for the oscillation in the uniform density 
with the Cowling approximation~\cite{koho99,prbero81}.

Present authors applied the method to the axial 
oscillation in a rotating relativistic star
with the Cowling approximation~\cite{koho99}, 
where the metric perturbations were neglected.
They took account of the rotational correction up to
third order to examine the oscillation equation.
They pointed out the importance of the 
frame-dragging effect, which causes
different property unlike the Newtonian case. 
Thus, the perturbative approach proves useful in providing a
physical understanding of many processes.
In this paper, we extend the approximation scheme to include 
the metric perturbations.
The remainder of the paper is organized as follows.
In Sec. II, we formulate the perturbation scheme to
solve the linearized Einstein equations.
We employ the slow rotation approximation, i.e.,
angular velocity is assumed to be small expansion parameter.
We look for the solution in which axial-led functions are dominant,
where the axial-led functions mean the functions 
describing axial modes in the absence of the rotation.
We use the terminology polar-led  in the same way.
In our scheme, the lowest order equations are determined 
only by the axial-led functions. They were already derived 
elsewhere~\cite{ko98,beko99,lofr99,ko97}
but are reviewed in Sec. III.
In Sec. IV,
first-order corrections to the polar-led functions are shown.
In Sec. V, second-order correction terms are added to the 
same equations in the form as the lowest ones.
In Sec. VI, concluding remarks are given.
Throughout this paper,
we work in the geometrical units of $c =G =1$.

\section{   Perturbation scheme  }

We consider a rotating star with uniform angular velocity 
 $\Omega \sim  O(\varepsilon), $
where 
$ \varepsilon $ is a small rotational parameter. 
The metric and fluid quantities describing the equilibrium state 
can be calculated by the slow rotation formalism~\cite{ha67,qu76}.
They are summarized in the Appendix A.
We next investigate the perturbations from the state.
The metric perturbations can be described by six functions.
Working in the Regge-Wheeler gauge~\cite{rewh57},
the perturbations are expressed as
\begin{equation}
h_{\mu\nu} =
\sum_{l,m}
\left(\begin{array}{cccc}
 e^{\nu}H_{0lm}(t,r)Y_{lm} 
 & H_{1lm}(t,r)Y_{lm} 
 & -h_{0lm}(t,r) \frac{\partial_{\phi}Y_{lm}}{\SA{}}
 & h_{0lm}(t,r)\SA{}\partial_{\theta}Y_{lm}\\
 \mbox{sym}
 & e^{\lambda}H_{2lm}(t,r)Y_{lm}
 & -h_{1lm}(t,r) \frac{\partial_{\phi}Y_{lm}}{\SA{}}
 & h_{1lm}(t,r)\SA{}\partial_{\theta}Y_{lm}\\
 \mbox{sym}     
 & \mbox{sym}     
 & r^2 K_{lm}(t,r)Y_{lm}
 & 0\\
 \mbox{sym} & \mbox{sym} & \mbox{sym} & r^{2}\SA{2}K_{lm}(t,r)Y_{lm}
\end{array}
\right),
\end{equation}
where
$Y_{lm} =Y_{lm}(\theta,\phi)$ represents spherical harmonics.
The symbol ``sym'' indicates that the missing components of $h_{\mu\nu}$
are to be found from the symmetry $h_{\mu\nu} =h_{\nu\mu}$.
The angular part is expanded with an appropriate combination 
of the harmonics.
In a same way,  the fluid perturbations are
described by five functions. They are
the pressure perturbation $\delta p$,
density perturbation $\delta \rho$
and three components of the 4-velocity  
($\delta u_r , \delta u_\theta ,  \delta u_\phi $).
The component $\delta u_{t}$ can be determined by the condition,
$\delta u_{\mu} u^{\mu} =0$.
These perturbed quantities are also expanded as
\begin{eqnarray}
 \delta p
&=&
\sum_{l,m}\delta p_{lm}(t,r)Y_{lm},
\\
 \delta\rho
&=&
\sum_{l,m}\delta \rho_{lm}(t,r)Y_{lm},
\\
 (\rho + p)\delta u_{r}
&=&
 e^{\nu/2} \sum_{l,m}R_{lm}(t,r) Y_{lm},
\\
 (\rho + p)\delta u_{\theta}
&=&
 e^{\nu/2}\sum_{l,m}
  \left[
   V_{lm}(t,r) \partial_{\theta} Y_{lm}
   -
   \frac{U_{lm}(t,r)}{ \SA{} }
   \partial_{\phi} Y_{lm}
   \right],
\\
 (\rho + p)\delta u_{\phi}
&=&
 e^{\nu/2}\sum_{l,m}
 \left[
  V_{lm}(t,r) \partial_{\phi} Y_{lm}
  +
  U_{lm}(t,r) \SA{} \partial_{\theta}Y_{lm}
  \right] .
\label{end-p}
\end{eqnarray}
These eleven functions are determined by ten components of
 the linearized Einstein field equations,
\begin{equation}
 \delta G_{\mu\nu}
=
8\pi \delta T_{\mu\nu},
\label{p-ein}
\end{equation}
and the adiabatic condition for the perturbations,
\begin{equation}
 \delta p +\xi\cdot\nabla p
=
\frac{\Gamma p}{p +\rho}(\delta \rho +\xi\cdot\nabla\rho),
\label{thermo}
\end{equation}
where
$\Gamma$ is the adiabatic index and $\xi$ is the Lagrange displacement.

Now we will solve the pulsation equations 
by the expansion of the spherical harmonics.
In the spherically symmetric star, the equations are decoupled
for each harmonic index $ (l,m).$
The perturbations can also be decoupled into
the axial and polar perturbations. They are respectively described
by the axial functions ${\cal A}_{lm}$
$ \equiv  (U_{lm},h_{0,lm},h_{1,lm}),$
and the polar functions  ${\cal P}_{lm}  $
$ \equiv ( \delta p_{lm},
 \delta \rho_{lm}, R_{lm},  V_{lm},$ 
$ H_{0,lm}, H_{1,lm}, H_{2,lm}, K_{lm}).$
In the presence of rotation, the perturbations
are described by the mixed state of them.
From now on, we call these functions
as the axial-led ones for ${\cal A}_{lm}$
and the polar-led ones for ${\cal P}_{lm}.$
Since the slow rotation is associated with
the perturbation with $ l=1, $
the formal relation  
between ${\cal A}_{lm}$
and ${\cal P}_{lm}$ in the Eqs.(\ref{p-ein})-(\ref{thermo})
can schematically be expressed as
\begin{eqnarray}
 0 =  [ {\cal A}_{lm} ] +
  {\cal E} \times  [ {\cal P}_{l\pm 1m} ]+
  {\cal E}^2 \times [ {\cal A}_{lm} ,  {\cal A} _{l\pm 2m} ]
  + \cdots,
\label{cpl1.eqn}
\\
 0 =  [ {\cal P}_{lm} ]+
  {\cal E} \times [ {\cal A}_{l\pm 1m} ]+
  {\cal E}^2 \times [ {\cal P}_{lm} ,  {\cal P} _{l\pm 2m} ]
  + \cdots ,
\label{cpl2.eqn}
\end{eqnarray}
where the symbol ${\cal E} $ denotes some functions of order
$ \varepsilon ,$  and
the square bracket formally represents the relation
among the perturbation functions therein.
These selection rules follow from the addition of 
angular momenta.
We moreover assume that the axial-led functions are dominant
in the slowly rotating star, i.e.,
$ {\cal A}_{lm}  \gg {\cal P}_{lm} .$
This assumption is not valid for the some cases. 
If the star and its perturbations obey the same one-parameter
equation of state, r-modes and g-modes are coupled at the 
lowest order in general.
They are respectively described by 
$ {\cal P}_{lm} $ and $ {\cal A}_{lm} ,$
which have zero frequency in the spherical isentropic star.
They form hybrid modes in the rotating star~\cite{lofr99}.
However, the r-modes are discriminated from the g-modes
for non-isentropic stars,
since the g-mode frequencies are non-zero in the spherical limit.
In that case, we may use the assumption 
$ {\cal A}_{lm}  \gg {\cal P}_{lm} $ 
to solve the r-mode oscillations, and expand as
\begin{equation}
    {\cal A}_{lm} =  {\cal A}_{lm} ^{(1)}+
 \varepsilon ^2 {\cal A}_{lm} ^{(2)}  + \cdots,
~~~
    {\cal P}_{lm} =  \varepsilon (
{\cal P}_{lm} ^{(1)}+  \varepsilon ^2 {\cal P}_{lm} ^{(2)} +
 \cdots ) .
\label{expand}
\end{equation}

Substituting these functions into
Eqs.(\ref{cpl1.eqn})-(\ref{cpl2.eqn}),
and comparing each order of $\varepsilon $,
we have the following equations of $ \varepsilon^n (n=0,1,2)$,
\begin{eqnarray}
0 &=&  [ {\cal A}_{lm} ^{(1)} ],
\label{pts.1}
\\
0 &=&  [ \varepsilon {\cal P}_{l\pm1m} ^{(1)}  +
{\cal E} \times  {\cal A}_{lm} ^{(1)} ],
\label{pts.2}
\\
0 &=&   [  \varepsilon ^2 {\cal A}_{lm} ^{(2)} ]    +
{\cal E} \times [ \varepsilon  {\cal P}_{l\pm 1m} ^{(1)}]
+ {\cal E}^2 \times
[ {\cal A}_{l m} ^{(1)},  {\cal A}_{l\pm 2m} ^{(1)} ]
\label{pts.3}
\\
&=& [ \varepsilon ^2 {\cal A}_{lm} ^{(2)}
 +
 {\cal E}^2 \times  {\cal A}_{l m} ^{(1)}] .
\label{pts.4}
\end{eqnarray}
We have here assumed that the perturbation in the lowest order
is described by a single component of spherical harmonic, 
that is,
$    {\cal A}_{l' m} ^{(1)}=0 $, for $ l' \neq l  $. 
The polar-led functions in Eq.(\ref{pts.3}) are eliminated by
 Eq.(\ref{pts.2}).
Equation (\ref{pts.1}) represents the axial oscillation in 
the lowest order.
Equation (\ref{pts.4}) is the second-order form of it, and
the term $ {\cal E}^2 \times {\cal A}_{l m} ^{(1)} $
can be regarded as the rotational corrections.
  The method to solve the equations is straightforward.
The first-order equations are solved by the axial-led
functions. The polar-led functions are expressed using them.
We have the second-order equations with the corrections
expressed by the axial-led functions in the lowest order.
These equations are successively solved in the following sections.
In the actual calculations, we also assume that
the time variation of the oscillation is slow and
proportional to $\Omega, $ i.e.,
$\partial _t \sim \Omega \sim  O(\varepsilon) .$
This is true in the r-mode frequency, 
$\partial _t \sim (1 -2/l(l+1) )m \Omega .$

\section{ Lowest-order calculation }

In this section, we review the equations governing the 
axial oscillations at the lowest order.
The radial functions are decoupled for each
spherical harmonic index. 
We denote it as $ L \equiv (l,m)$ for the abbreviation.
The relevant functions are
$ h_{0,L} , h_{1,L} $  and $U_{L}.$ 
They are calculated from three components,
essentially  $(t \phi)$ $(r \phi)$ and $(\theta \phi)$ components,
of the Einstein equations.
We define a function $ \Phi^{(1)} _{L} $ as 
\begin{equation} 
  \Phi^{(1)}_{L} =\frac{h^{(1)}_{0,L}}{r^2},
\end{equation}
where the superscript $^{(1)}$ denotes the lowest order term in
Eq.(\ref{expand}).
The relation between the metric functions is given by
\begin{equation}
 h^{(1)}_{1,L}
=
\frac{r^4 e^{-\nu}}{[\LL -2]}
\left[
(\partial_{T} -im\varpi)\Phi^{(1) \prime}_{L}
+\frac{2im\varpi'}{\LL}\Phi^{(1)}_{L}
\right],
\label{h1-Phi}
\end{equation}
where  $\partial_{T}=\partial_{t} +im\Omega$
denotes a time-derivative in a co-rotating frame and
a prime denotes a derivation with respect to $r.$
The axial velocity function is expressed by two ways: 
\begin{eqnarray}
 (\partial_{T} -im\chi)U^{(1)}_{L}
&=&
-4\pi(p_{0} +\rho_{0})r^2e^{-\nu}\partial_{T}\Phi^{(1)}_{L},
\label{U-Phi}
\\
U^{(1)}_{L}
&=&
\frac{j^2r^2}{4}\left[
		 \frac{1}{jr^4}(jr^4\Phi^{(1) \prime}_{L})'
		 -(v +16\pi(p_{0} +\rho_{0})e^{\lambda})\Phi^{(1)}_{L}
		\right],
\label{U-ddPhi}
\end{eqnarray}
where  
\begin{eqnarray}
 \chi
&=&
\frac{2}{l(l+1)}\varpi
=
\frac{2}{l(l+1)}(\Omega -\omega),
\\
  v &=&  \frac{e^{ \lambda }}{r^2} \left[
       l(l+1) -2 \right] , 
\\
 j &=& e^{- (\lambda +\nu )/2} .
\end{eqnarray}
Eliminating $U^{(1)} _{L} $ in Eqs.(\ref{U-Phi})-(\ref{U-ddPhi}),
we have a second-order differential equation for $\Phi^{(1)}_{L}$,
\begin{eqnarray}
0 &=& {\cal  L } [ \Phi^{(1)}_{L}]
\nonumber
\\
 &=&
 \left( \partial _T  - i m \chi \right)
 \left[ \frac{1}{j r^4}
 \left( j r^4   \Phi^{(1) \prime} _{L}  \right)' 
 - v  \Phi^{(1)} _{L}  \right]   
    + 16 \pi i m  \chi (p_{0} +\rho_{0}) e^{ \lambda } 
 \Phi^{(1)} _{L} .
\label{master-Phi1}
\end{eqnarray}
The linear operator ${\cal L}$ is defined above.
This equation is reduced to a singular eigen-value 
problem~\cite{ko98}, 
by assuming the time dependence for the mode as the form
$ e^{-i\sigma t } .$
The equation can be written as 
\begin{equation}
 (\varpi -\mu)
\left[
\frac{1}{jr^4}(jr^4 \Phi^{(1) \prime}_{L})'
-v\Phi^{(1)}_{L}
\right]
=
 q\Phi^{(1)}_{L},
\label{eqn.ray}
\end{equation}
where 
\begin{eqnarray} 
  \mu &=&
 - \frac{ l(l+1) }{2m}(  \sigma -m \Omega ),
\\
  q & = & \frac{1}{j r^4} \left( j r^4  \varpi ' \right)' 
    = 16\pi (p_{0} +\rho_{0})e^{\lambda}\varpi \geq 0 .
\end{eqnarray}
There is a singular point $r_0 $  in Eq.(\ref{eqn.ray}) 
unless $q(r_0) = 0,$ corresponding to
the real value of $ \mu =  \varpi (r_0) .$
It is evident that the singularity originates from
the mismatch in Eq.(\ref{master-Phi1}), i.e.,
the first term vanishes whereas the second never does.
When the first term, which is formally of first order,
is small enough, then higher order corrections  become important.
This situation is very similar to the inviscid shear flows.
When the viscosity is small enough, the stability is almost 
determined by the Rayleigh equation, the perturbation equation 
for the inviscid theory.
The Rayleigh equation has critical points for some mean fluid.
The viscous corrections should be included to determine
the behavior near the neighborhood of the critical points.
Therefore, the equation should be replaced by the Orr-Sommerfeld 
equation derived from the Navier-Stokes equation.

As we will show in the subsequent sections,
the function $\Phi^{(1)}_{L}$ can also affect
the polar-led functions.
When one considers the equations of the next order,
$\Phi^{(2)}_{L}$, 
additional terms appear in the form (\ref{master-Phi1}).
The terms depend on different aspect of the 
background flow, and play an important role as
the viscosity terms.

\section{First-order corrections in polar-led functions}

In this section, we will derive the equations governing 
the polar-led functions,
$H^{(1)}_{0}$, $H^{(1)}_{1}$,
$H^{(1)}_{2}$, $K^{(1)}$,
$\delta p^{(1)}$, $\delta \rho^{(1)}$, $R^{(1)}$ and
$V^{(1)}$.
As shown in Sec. II, the functions with $ (l\pm 1,m) $
are coupled with the axial-led functions with  $ (l,m) .$
We will shorten the subscript of the spherical harmonic index
as e.g., 
$\delta p^{(1)}_{\pm} \equiv  \delta p^{(1)}_{l\pm 1 m}$. 
These eight functions are determined by seven components of
the linearized Einstein field equations and one thermodynamical 
relation. 
The calculations are straightforward, but results are sometimes
messy.
The pressure $\delta p^{(1)}_{\pm}$ 
and density perturbations $\delta \rho^{(1)}_{\pm}$ 
are expressed by 
\begin{eqnarray}
4\pi\delta p^{(1)}_{\pm}
&=&
2\pi(p_{0} +\rho_{0})
(H^{(1)}_{0,\pm} +2 T_{\pm} \Omega r^2 e^{-\nu}\Phi^{(1)}_{L})
+2S_{\pm}\varpi U^{(1)}_{L},
\label{eqn.pres}
\\
4\pi \delta \rho^{(1)}_{\pm}
&=&
-4\pi\frac{\rho'_{0}}{\nu'}
(H^{(1)}_{0,\pm} +2 T_{\pm} \Omega r^2 e^{-\nu}\Phi^{(1)}_{L})
\NN
&&
-4S_{\pm}\frac{e^{-\nu/2}}{\nu'}
\left(  e^{\nu/2}\varpi U^{(1)}_{L}
\right)'
+2 T_{\pm}\frac{e^{\nu}}{\nu' r^2}
(\varpi r^2 e^{-\nu})'U^{(1)}_{L},
\label{eqn.dns}
\end{eqnarray}
where
\begin{equation}
 S_{+} = \frac{l}{l+1}Q_{+},
~~
S_{-}  = \frac{l+1}{l}Q_{-},
~~
T_{+} =  lQ_{+},
~~
T_{-} = -(l+1)Q_{-},
\end{equation}
\begin{equation}
Q_{+} = \sqrt{\frac{(l+1)^2 -m^2}{(2l +1)(2l +3)}},
~~
Q_{-} = \sqrt{\frac{l^2 -m^2}{(2l -1)(2l +1)}}.
\end{equation} 
These quantities are expressed by the axial-led functions 
($\Phi^{(1)}_{L},U^{(1)}_{L})$ and 
$H^{(1)}_{0,\pm}. $
There is another relation among  $H^{(1)}_{0\pm} ,$
$K^{(1)}_{\pm} $
and  $\delta \rho^{(1)}_{\pm}$ in the field equations. 
Eliminating $\delta \rho^{(1)}_{\pm}$ by Eq.(\ref{eqn.dns}), 
we have a second-order differential equation for the
metric perturbations. The equation can be regarded as
relativistic version of the Poisson equation,
$ \nabla ^2 \delta \phi = 4 \pi \delta \rho ,$
for the gravitational potential $ \delta \phi $ and
the density perturbation $\delta \rho .$
In the relativistic case,  
the Newtonian potential is replaced by
$H^{(1)}_{0}$ or  $K^{(1)}.$ 
The second-order differential equation 
is explicitly given by
\begin{eqnarray}
& &K^{(1) \prime} _{\pm} 
-\frac{e^{\lambda}}{\nu' r^2}\{
n_\pm (K^{(1)}_{\pm} -H^{(1)}_{0,\pm})
+
2( 4\pi(p_{0} +\rho_{0}) r^2 
+e^{-\lambda} -1)H^{(1)}_{0,\pm}
\}
= s_{1},
\label{eqn.kdf}
\\
& &(K^{(1)}_{\pm} - H^{(1)}_{0,\pm}  )' 
-\nu'H^{(1)}_{0,\pm}
= \left(1 + \frac{\nu' r}{2}\right) s_{1}+s_{2},
\label{eqn.khdf}
\end{eqnarray}
where
\begin{eqnarray}
n_{\pm } &=&
-2 +l'(l'+1)|_{ l'= l\pm 1},
\\
s_{1}
 &=&
- Q_{\pm}\left[
	8\varpi r U^{(1)}_{L}
	+32\pi(p_{0} +\rho_{0})\varpi r^3 e^{-\nu} \Phi^{(1)}_{L}
	+2 \varpi' j^2 r^3 \Phi^{(1) \prime}_{L}
       \right]
\NN
&&
+S_{\pm}\Biggl[
	 8\left\{r +\frac{e^{\lambda}}{\nu'}\right\}\varpi U^{(1)}_{L}
	 -\frac{4 e^{-\nu}}{\nu'} \left\{
            l(l+1) \omega e^{\lambda }
            +16\pi p'_{0} \varpi r^3 
	    -\varpi' r 
       \right\}\Phi^{(1)}_{L}
       +2 \varpi' j^2 r^3 \Phi^{(1) \prime}_{L}
	\Biggr]
\NN
&&
+T_{\pm}\left[
	 \frac{2 e^{-\nu}}{ \nu'}\left\{
	   (2 -2e^{\lambda} -l(l+1) e^{\lambda})\omega
	   +8\pi \Omega (p_{0} +\rho_{0}) r^2 e^{\lambda}
	   -\varpi' r
       \right\}
\Phi^{(1)}_{L}
       -\frac{\varpi'r^2 e^{-\nu}}{ \nu'} \Phi^{(1)'}_{L}       
	\right],
\NN
\\
s_{2}
&=&
\left( r+ \frac{2}{\nu'} \right)
\Biggl[ -S_{\pm}
     \left [ 4\varpi e^{\lambda} U^{(1)}_{L}
  -2\left\{l(l+1) \omega e^{\lambda }
   - 2\varpi'r  \right\}  e^{ -\nu}\Phi^{(1)}_{L}
  \right ]
\NN
&&
+T_{\pm} \left\{
\frac{1}{2} \varpi' r^2 e^{-\nu} \Phi^{(1) \prime}_{L} +
\left[ \left\{ (l(l+1)+2)\omega -8\pi(p_{0} +\rho_{0})\Omega r^2 
 \right\} e^{\lambda -\nu}
 +2 \omega e^{ -\nu }
       \right]   \Phi^{(1)}_{L} \right\} 
\Biggr]
\NN
&&
+
\Biggl[ 4 S_{\pm} \frac{ \varpi' r e^{-\nu}}{\nu'}\Phi^{(1)}_{L}
+ T_{\pm}\left( 2\omega r^2 e^{-\nu} \Phi^{(1) \prime}_{L}  
+\frac{ (2 \varpi' r - 8 \omega)e^{-\nu}}{ \nu'}\Phi^{(1)}_{L}\right)
	\Biggr].
\end{eqnarray}
When the axial-led function 
at the lowest order $\Phi^{(1)}_{L}$ 
is given, the functions $H^{(1)}_{0,\pm}$ and $K^{(1)}_{\pm}$
are solved with appropriate boundary conditions.
In a similar way, we can solve the other polar-led functions,
$H^{(1)}_{1,\pm}$, $H^{(1)}_{2,\pm}$, $R^{(1)}_{\pm}$ and 
$V^{(1)}_{\pm}$
by ($\Phi^{(1)}_{L},U^{(1)}_{L})$ and 
$(H^{(1)}_{0,\pm},K^{(1)}_{\pm}).$
The expressions for these four polar-led functions are omitted here,
since they are eliminated in the following calculations, and 
never appear in the final results.
However, here is a comment on using the adiabatic 
condition Eq.(\ref{thermo}). The time derivative 
of it can be written as
\begin{eqnarray}
4\pi \partial_{T} \left (
\delta p^{(1)} _{\pm}
  - \frac{\Gamma p_{0}}{p_{0} +\rho_{0}}
\delta \rho^{(1)} _{\pm} \right )
&=&
\frac{A \Gamma p_{0} e^{\nu} }{p_{0} +\rho_{0}}
\left(
	   e^{-\lambda}R^{(1)} _{\pm}
	   -\frac{3im\xi_{2}}{r^2}Q_{\pm}U^{(1)} _{L}
	   \right),
\end{eqnarray}
where
\begin{eqnarray}
\xi_{2}
&=&
-\frac{2}{\nu'}
\left(
 h_{2} +\frac{1}{3} \varpi^2 r^2 e^{-\nu}
\right),
\\
A
&=&
\frac{\rho'_{0}}{p_{0} +\rho_{0}}
-\frac{p'_{0}}{\Gamma p_{0}}.
\end{eqnarray}
This thermodynamical relation determines
the function $R^{(1)} _{\pm}$ unless
the Schwarzschild discriminant $A$ vanishes.
Otherwise, we have 
one constraint for the function  $U^{(1)} _{L}$
through Eqs.(\ref{eqn.pres})-(\ref{eqn.dns}), and
the function $R^{(1)} _{\pm}$ should be specified
in another way.
(For example, see the method by~\cite{koho99} 
in the Cowling approximation.)
The mathematical drawback for the isentropic case $ A= 0$ 
is related with the coupling of the g-modes.
Both r-modes and g-modes are degenerate to zero frequency
in the non-rotating star, and hence 
a particular treatment is necessary~\cite{lofr99}.
From now on,  we will consider the case  $ A \ne 0 $ only.

\section{Including second-order corrections}

So far we have considered ten components of the Einstein equations
and one thermodynamical relation. They were limited to the
lowest-order form with respect to the rotational parameter.
In this section, we consider how the next order terms
modify the equation (\ref{master-Phi1}) derived in Sec. III.
The relevant equations for this purpose are
three components of the Einstein equations, 
i.e., $(t \phi)$ $(r \phi)$ and $(\theta \phi)$ components,
which are used in the leading order equation in  Sec. III.
These equations contain the relations among
the axial-led functions of second-order
$ h^{(2)}_{0,L}$, $h^{(2)}_{1,L}$, $U^{(2)} _{L}$
and the polar-led functions calculated in the previous section.
We follow the same procedure as done in Sec. III.
Defining the function $\Phi^{(2)}_{L} =  h^{(2)}_{0,L}/r^2 $
and eliminating $ h^{(2)}_{1,L} $ and $U^{(2)} _{L},$
we eventually have the equation governing the axial oscillations.
It can be written in the following form
\begin{eqnarray}
{\cal L}[\Phi^{(2)}_{L}] = 
{\cal D}[\Phi^{(1)}_{L}, h^{(1)}_{1,L}, U^{(1)}_{L}]
+{\cal G}[H^{(1)}_{0,\pm}, K^{(1)}_{\pm}],
\label{master-Phi}
\end{eqnarray}
where the operator ${\cal L} $ is defined in Eq.(\ref{master-Phi1})
and the right hand side means the second-order rotational 
corrections. They consist of several terms as
\begin{eqnarray}
{\cal D}[\Phi^{(1)}_{L}, h^{(1)}_{1,L}, U^{(1)}_{L}]
&=&
{\cal D}_{0}
+\alpha_{1} h^{(1)}_{1, L}
+\alpha _{2} \partial _t \Phi ^{(1)}_{L}
+im ( \beta_{1} U^{(1)}_{L}
+ \beta_{2}\Phi'^{(1)}_{L}
+ \beta _{3} \Phi^{(1)}_{L} ),
\end{eqnarray}
\begin{eqnarray}
{\cal G}[H^{(1)}_{0,\pm}, K^{(1)}_{\pm}]
&=&
 ( A_{1}\partial_{T} +im A_{2} )K^{(1)}_{\pm}
+( B_{1}\partial_{T} +im B_{2} )H^{(1)}_{0,\pm}.
\end{eqnarray}
The explicit forms of the coefficients $\alpha_{i}$, 
$\beta_{i}$, $A_{i}$ and  $B_{i}$ are given in Appendix B.
They are expressed by the quantities determined 
by the stellar model in the equilibrium.
Since the term ${\cal D}_{0}$ contains
higher order derivatives,
we explicitly show below:
\begin{eqnarray}
 {\cal D}_{0}
&=&
 32 c_3 \frac{ \varpi e^{ -\nu} }{j r^2}
    \left \{
     \frac{ (\rho_0 + p_0) r^2}{ j A \nu '}
        \left( \frac{ \varpi \partial _T  U^{(1)} _{L} }
          { \rho_0 + p_0 }
        \right)'
      \right \}' 
\NN
&&
 + \left \{ 
   \frac{16 c_2 \varpi^2 e^{-\nu}}{(\rho_0 + p_0)j r^2}
    \left[
    \frac{ (\rho_0 + p_0) e^{\nu}}{A \varpi j \nu'}
    (\varpi r^2 e^{-\nu})'
    \right]'
  -  \left(
    \frac{8 c_1 e^{\nu}}{A j^2 \nu'r^4}
    \right)
   [( \varpi r^2 e^{-\nu})']^2
\right  \} \partial _T  U^{(1)} _{L}
\NN
&&
+( \partial _T -im \chi )
\left \{
\left( \frac{\nu'}{r^2} -\frac{2}{r^3} \right)
 ( \partial _T  -im\varpi  )
+ 2 im   \frac{\varpi'}{r^2} 
\right \}
h^{(1)}_{1,L}
\NN
&&
-e^{\lambda -\nu} 
( \partial _T -im\chi  )
( \partial _T -im\varpi )^2
\Phi^{(1)}_{L},
\label{df4th}
\end{eqnarray}
where
\begin{equation}
 c_{n}  =
    \frac{l+1}{l^n} Q^2_{-}
   +(-1)^{n-1}
   \frac{l}{(l+1)^n} Q^2_{+}  .
\end{equation}
From Eqs.(\ref{master-Phi1}) and (\ref{master-Phi}),
a function $
 \Phi_{L} =  \Phi^{(1)}_{L}  + \varepsilon^2 \Phi^{(2)}_{L} 
$ satisfies the following equation, which is correct up to
$O(\varepsilon ^2 ),$ 
\begin{eqnarray}
{\cal L}[\Phi_{L}] = 
{\cal D}[\Phi_{L}, h_{1,L}, U_{L}]
+{\cal G}[H_{0,\pm}, K_{\pm}].
\label{mod-master-Phi}
\end{eqnarray}
The quantities without the superscript satisfy the same
relations as in the leading order, i.e.,
Eqs.(\ref{h1-Phi}), (\ref{U-ddPhi}), (\ref{eqn.kdf})
and (\ref{eqn.khdf}), which are adequate approximation to 
this order.
This equation is of course reduced to the 
leading order equation (\ref{master-Phi1}), when
the second-order rotational effects 
and the coupling to the polar modes are neglected.

The first term in ${\cal D}_{0}$
contains fourth derivative of $\Phi_{L}$
with respect to $r$, 
since  $U _{L}$ can be expressed by the
second derivative of $\Phi_{L}$
as shown in Eq.(\ref{U-ddPhi}).
The highest derivative  term of $\Phi _{L}$
with respect to $ r $ is therefore given by
\begin{equation}
 {\cal D}_* [\Phi_{L}] =
8 c_3 \frac{ \varpi e^{ -\nu} }{j r^2}
    \left \{
     \frac{ (\rho_0 + p_0) r^2}{ j A \nu '}
        \left( \frac{ j \varpi }{( \rho_0 + p_0 )r^2 }
 ( j r^4 \partial _T  \Phi _{L} ' )'
        \right)'
      \right \}' .
\label{dstar} 
\end{equation}
Neglecting $ {\cal G} $ and  $ {\cal D} $ except
$ {\cal D}_* $ in Eq.(\ref{mod-master-Phi}) leads to 
$ {\cal L}[\Phi_{L}] = {\cal D}_* [\Phi_{L}], $
which is analogous to the Orr-Sommerfeld equation in 
the incompressible shear flow. (See Ref.~\cite{ko99}.)
The term (\ref{dstar}) effectively gives the 'viscosity' in
the viscous fluid.
The viscosity is important for the stability of the flows.
For the small Reynolds number, the laminar flow is realized,
whereas the flow becomes turbulence above 
a critical Reynolds number.
The effective Reynolds number $ R_e $ in Eq.(\ref{dstar})
is estimated from dimensional argument as 
\begin{equation}
  R_e \sim \frac{A \nu '}{ \varpi ^2} .
\end{equation} 
This is roughly the square of the ratio of
the g-mode frequency to rotational one. 
The viscosity term will play
a key role on the singular point of the first-order
equation, but the consequence is not clear at moment.
It is necessary to explore further how the effective 
Reynolds number should operate in the stability and so on.

\section{Concluding remarks}

In this paper, we have explored an effective theory
to describe the axial oscillation on a slowly rotating stars. 
The approximate equation governing the oscillation 
is constructed from the Einstein equations.
The equation is derived by assuming 
that the angular dependence of the oscillation is 
dominated by a single component of spherical harmonics.
The assumption in general breaks down as the evolution of
oscillation.  There are coupling terms
of order $\varepsilon  $  in the rotating fluids. 
Other oscillation patterns with different spherical harmonics
will gradually be produced through the rotational coupling.
For this reason, the equation is valid for small $t$, and 
can be used to examine the time evolution as
the initial-value problem.
The rotational effects up to third order are involved in this
paper, so that the regime of application is enlarged.
The equation derived here is also irrelevant to
the singular point found in the first order one.

   The equation also shows a remarkable property.
It is evident that the axial oscillation strongly
couples to g-mode oscillations.
Viscosity-like term arises from the polar pieces 
related to the g-mode oscillations.
The 'viscosity' term originated from considering the 
sub-system only, i.e., a single component of the spherical harmonics. 
There are no production and extinction in the whole system, 
but, e.g., 'energy' of a component is partially transfered to 
the others. This transportation is regarded as dissipative effect 
so far as a particular sub-system is concerned.
The term also has a significant implication.
The condition $A=0 $ is a good approximation for 
cold neutron stars, so that the coupling may be neglected.
On the other hand, it is not clear that the condition holds, 
in particular for newly born hot neutron stars, in which
the r-mode instability sets in.

In this paper, we concentrated the equations only inside a star,
to be more precise, the equations for the region $A \ne 0 .$ 
The pulsation equation derived here should be solved with 
appropriate boundary conditions. The boundary conditions
are determined from matching with the equations outside, or
regularity conditions. 
For example, the regularity condition of a function $ \Phi  $ 
is given by $ \Phi  \sim r^{l-1} $ near the center.
Depending on details of the stellar structure, 
the solution for the  $A \ne 0 $ region  may be matched to
the solution for isentropic region, $A=0$. 
Furthermore, the interior solution should be matched to the 
exterior one at the stellar surface.
The exterior perturbation equation in vacuum is not derived here, 
but the form  should be reduced to the wave equation describing 
gravitational wave.  The perturbation equation should be solved by 
out-going boundary condition at infinity. 
One question may arise.  Is it possible to calculate the radiation 
reaction at this order?  Newtonian estimate indicates that 
the back-reaction is of order 
$\varepsilon^{2m +2} $ for $m \ge 2 .$
Our expansion of the rotational parameter is limited to the 
third-order, and higher order corrections are necessary to examine 
the effect in a consistent way.  
The radiation-reaction effect is also a kind of dissipative one, 
so that the accurate evolution for a long period is necessary.

\section*{Acknowledgements}
We would like to thank John Friedman and Nils Andersson 
for helpful discussions. This was supported in part 
by the Grant-in-Aid for Scientific Research Fund of
the Ministry of Education, Science and Culture of Japan(08NP0801).

\appendix

\section{Equilibrium configuration of a slowly rotating perfect fluid}

We here summarize the equilibrium of a slowly rotating 
star to explain our notation.
The equilibrium state with uniform angular velocity
$\Omega \sim O(\varepsilon)$
can be described by
stationary and axisymmetric metric $g_{\mu\nu}$,
4-velocity $u^{\mu} =(u^{t}, 0, 0, u^{\phi})$,
pressure $p$ and energy density $\rho$ of the fluid.
The rotational corrections up to $O(\varepsilon^3)$ is 
needed to assure the consistency in our analysis. 
The metric is given by
\begin{eqnarray}
 ds^2
&=&
-e^{\nu}[1 +2(h_{0} +h_{2}P_{2})]dt^2
+e^{\lambda}[1 +\frac{2 e^{\lambda}}{r}(m_{0} +m_{2}P_{2})]dr^2
\NN
&&
+r^2(1 +2k_{2}P_{2})
\left\{
 d\theta^2 +\SA{2}\left[
		   d\phi -\left(
			    \omega +W_{1} 
			    -W_{3}\frac{1}{\SA{}}\frac{dP_{3}}{d\theta}
			    \right)dt
		  \right]^2
\right\},
\end{eqnarray}
where $P_{l}=P_{l}(\CO{})~(l=2,3)$ is the Legendre polynomial of order $l$.
The metric functions introduced above obey the following ordering in
$\varepsilon$:
$\omega \sim O(\varepsilon)$,
$h_{0}, h_{2}, m_{0}, m_{2}, k_{2} \sim O(\varepsilon^2)$,
$W_{1}, W_{3} \sim O(\varepsilon^3)$.
These are functions of radial coordinate $r$ only.
The components of the 4-velocity 
are
\begin{equation}
 u^{t}
=
(-g_{tt} -2g_{t\phi}\Omega -g_{\phi\phi}\Omega^2)^{-1/2},
~~
u^{\phi}
=
\Omega u^{t}.
\label{eq.of.4-velo}
\end{equation}
The pressure and energy density are respectively given by
\begin{equation}
  p
=
p_{0} +\{p_{20} +p_{22}P_{2}\},
~~
\rho
=
\rho_{0} +\{\rho_{20} +\rho_{22}P_{2}\},
\label{eq.of.p2d2}
\end{equation}
where
$p_{0}$ and $\rho_{0}$ are the pressure and energy density
of non-rotating fluid. 
The centrifugal force of order $\varepsilon^2$ alters the
configuration shape, which corresponds to 
the quantities in the braces. 
The functions $p_{20}, p_{22}, \rho_{20}$ and $\rho_{22} $
are related with the metric functions of order  $\varepsilon^2$.
We rather use the metric functions to eliminate
the pressure and density of order $\varepsilon^2$ 
in the oscillation equations.

\section{The second-order terms}

\subsection{Coefficient of $ h^{(1)}_{1,L}$, $\alpha_{1}$}
\begin{eqnarray}
 \alpha_1 &=&
c_2 \Biggl[
 32\pi(\rho_{0} +p_{0} )
\left(
 \frac{2}{ \nu'^2  j^2 r^2 } \varpi (\omega e^{-\nu})'
 +\frac{2}{r} \varpi \omega 
 - (\omega+ 3 \varpi) \varpi' 
\right )
- \frac{8}{ r^2 e^{\lambda} }(\omega r)'\varpi'
\Biggr]
\NN
&+&c_1 \Biggl[
-5\pi(\rho'_{0} +3 p'_{0}) \varpi^2 
+\left(
 \frac{28}{3r^2e^{\lambda}} 
 -\frac{5}{12r^2}
 +\frac{5}{2r^3\nu'}
\right)\omega\varpi'
-\frac{97}{12re^{\lambda}}\varpi'^2
\NN
&&
+4\pi(p_{0} +\rho_{0})
\left(
 - \frac{5 \varpi^2}{r}
   +23\varpi\varpi'
  +\frac{28}{3}\omega\varpi' 
  -\frac{56}{3r}\omega\varpi
  -\frac{12 }{j^2 r^2\nu'^2} \varpi (  \omega  e^{-\nu})'
 \right)
\Biggr]
\NN
&+ &\frac{m^2}{l(l+1)} \Biggl[
2\pi \left( l(l+1)-\frac{11}{2} \right )
  ( \rho'_{0}+3p'_{0} ) \varpi^2
+\frac{4\pi}{r^2}\left( l(l+1)-\frac{15}{2} \right )
  (\rho_{0}+p_{0}) (\varpi^2 r^2)' 
\NN
&&
-\frac{ e^{\nu} }{2\nu'} \left( l(l+1)+\frac{3}{2} \right )
(e^{-\nu} r^{-2})' \omega \varpi'
- \frac{1}{2 r e^{\lambda}} \left( l(l+1)-\frac{1}{6} \right )
 \varpi'^2 
 +\frac{1}{3 r^2}( 8e^{-\lambda} -7) \omega\varpi'
\NN
&&
+16\pi(\rho_{0}+p_{0})\left(
 \frac{\varpi^2}{ r} 
 -\frac{ e^{\lambda }}{ r^2 \nu'^2}  \varpi \varpi'
 -\frac{e^{\lambda }}{\nu' r^2} \omega \varpi 
 +  \frac{2}{3} \omega r^2 ( \varpi r^{-2} )'  
 +3 \varpi \varpi'
\right )
\Biggr]
\NN 
&+&
\frac{4 }{r^3}e^{(6\nu  - \lambda)/4}
( e^{(-2 \nu  - 3 \lambda)/4} h_0' )'
-\frac{ e^{\nu} }{r^4}
(   l(l+1)-4 
   -( (\nu' r )^2- 6 ) e^{-\lambda}
 ) h_0'
\NN
&&
- \frac{1}{j^2 r^9} \left( 
\nu ' r^7 e^{- 2\lambda} \left( \frac{ e^{\lambda}}{r} m_0  \right)'
 \right)'
  -\frac{ e^{\nu} }{r^4}( l(l+1)- 2+4 e^{- \lambda}) 
   \left( \frac{ e^{\lambda}}{r} m_0  \right)'
\NN
&&
   -\frac{2\pi}{3}  (\rho'_{0}+3p'_{0} )
 \left\{ \left[ l(l+1) -4\frac{(e^{\nu} r^2)'}{r e^{\nu}} 
        \right]\varpi^2 
   - \frac{12 \nu' }{j^2 r^2}m_0 
 \right\}
\NN
&&
+[ 
 4\pi(\rho_{0}+p_{0}) r^2 (2 (\nu' r)^2
+\nu' \lambda' r^2+ 12 \nu' r + 2 \lambda' r +4 )
+8-2\lambda' r ]
 \frac{m_0}{j^2 r^6}
\NN
&&
+\frac{32\pi^2}{3}(\rho_{0}+p_{0})^2 \varpi^2 
e^{\lambda-\nu}( r^2 e^{\nu} )'
-\frac{16\pi}{3 r^5} e^{\nu /2 } 
(  r^6 \varpi ^2  e^{-\nu /2 } )' p'_{0}
\NN
&&
-\frac{8\pi}{3r}(\rho_{0}+p_{0}) [
  l(l+1) \varpi  ( \varpi r )' 
  -8 \omega  \varpi  +4 r \Omega \varpi'
 -2( 3 -  e^{\lambda} ) \varpi ^2 
]
\NN
&&
- \left \{
\frac{l(l+1)}{2 r^2} 
\left( 1+\frac{2}{ \nu'r} \right) 
+\frac{4}{3r^2 }(2e^{\lambda} -1)
\right \} \omega \varpi'
+\frac{1}{6 r  e^{\lambda}}
\left \{ l(l+1) -4 -3 \nu' r 
\right \} \varpi'^2.
\end{eqnarray}

\subsection{Coefficient of $ \partial_{t} \Phi^{(1)}_{L}$, $ \alpha_2 $}
\begin{eqnarray}
 \alpha_2 &=&
c_{2}\Biggl[
\frac{32\pi e^{\lambda-\nu}}{  \nu' } 
(\varpi +\Omega)\varpi r^2(\rho'_{0} +3p'_{0})
-32\pi(p_{0} +\rho_{0}) \frac{e^{\lambda-\nu}}{ ( r \nu')^2 }
 \{  (3 \varpi + \chi)
 ( 4 r^2 e^{\lambda} \varpi
\NN
&&
 - 2r \varpi e^{\nu}(r^2 e^{-\nu} )'
  - r^2 (r^2 e^{\nu})'( \varpi e^{-\nu} )'
 + 4r^2 e^{\lambda }\omega \varpi
- \nu'(r^4 \omega\varpi' +2(r^4\varpi^2)' )
\}
-\frac{4}{r \nu'}( r^2 e^{-\nu} )'\omega\varpi'
\Biggr]
\NN
& + &
c_{1}\Biggl[
16\pi(\rho'_{0} +3p'_{0})
\Omega (\varpi +\Omega)\frac{r^2 e^{\lambda -\nu}}{\nu'}
\NN
&&
-16\pi (p_{0} +\rho_{0}) \frac{e^{\lambda-\nu}}{  \nu'^2 }
\{
4 e^{\lambda }\omega\varpi
-2r \nu'(\varpi +\Omega )^2
- r^2 \nu' \varpi'( \varpi+ 3\Omega )
\}
\NN
&&
+ \frac{8e^{-\nu}}{r \nu'}\omega^2
 -2(r +\frac{2}{\nu'})e^{-\nu}\omega\varpi'
 -\frac{49}{24} r^2 e^{-\nu}\varpi'^2
-2f_1 +f_2  +f_4  
\Biggr]
\NN
& + &
\frac{m^2}{l(l+1)}\Biggl[
 l(l+1)
\left(\frac{2}{r \nu'}-1\right)e^{\lambda -\nu}\omega^2
-16\pi (p_{0} +\rho_{0}) r^2 e^{\lambda -\nu} \varpi^2
-\frac{71}{24}r^2 e^{-\nu}\varpi'^2
\NN
&&
-f_1 +f_2 - 3f_3-f_4
\Biggr]
\NN
& +&
l(l+1)\left(
-\frac{2e^{2\lambda}}{r^3} m_{0}
+\left(1 -\frac{2}{r \nu'}\right)e^{\lambda -\nu}\omega^2  \right)
-\nu'\left( \left(\frac{e^{\lambda} m_{0}}{r} \right)' +h_{0}'\right)
+\frac{4e^{2\lambda} m_{0}}{r^3}
\NN
&&
-16\pi p_{0}'e^{\lambda}
\left(e^{\lambda}m_{0} +\frac{r^3}{3}e^{-\nu}\varpi^2\right)
+2 r^2 e^{-\nu}\varpi'^2
+f_1-\frac{2}{3}f_2+ f_3,
\end{eqnarray}

where
\begin{eqnarray}
f_1 
&=& 
\frac{8\pi e^{\lambda}}{3}(p_{0} +\rho_{0})
\Biggl(
\biggl\{8e^{-\nu}r^2
-12r^2j^2
+\frac{64 e^{-\nu}}{ (\nu')^2}(4\pi r^2 e^{\lambda}p_{0} -r\nu')
\biggr\}\varpi^2
\NN
&&
+\frac{8 r^2}{\nu'}e^{-\nu}\left(1 +\frac{2}{\nu' r}\right)\varpi\varpi'
+r^{12}j^{2}\left(\frac{\varpi}{r^8}\right)'\varpi'
\Biggr)
+\frac{4 r}{3}e^{-\nu}\omega\varpi',
\\
f_2 
&=& 
4\pi(\rho'_{0} +3p'_{0})(
3e^{\lambda}\xi_{2} +2 r^3e^{-\nu}\varpi^2 ) 
-16\pi(p_{0} +\rho_{0})^2 r^4 e^{\lambda -\nu}\varpi^2
-12\left(1 - \frac{2}{r \nu'}\right) \frac{e^{\lambda}}{r^2}k_{2}
\NN
&&
+\frac{6}{r^3}\left( 
1  -3e^{\lambda} +4\pi(\rho_{0} +p_{0}) r^2e^{\lambda}
\right)\xi_{2}
+\frac{4}{r \nu'}( 1-3 e^{\lambda} ) e^{-\nu}\varpi^2
+ \left(\frac{r}{\nu'} - r^2 e^{-\lambda} \right) e^{-\nu}\varpi'^2,
\NN
\\
f_3 
&=& 
l(l+1)\left(
\frac{e^{\lambda}}{2r^2}\left( k_{2} -\frac{ \nu'}{2} \xi_{2}\right)
-\frac{4\pi}{3}(p_{0} +\rho_{0})r^2 e^{\lambda -\nu}\varpi^2
-\frac{e^{\lambda -\nu}}{6}\varpi^2
-\frac{r^2}{12}e^{-\nu}\varpi'^2 \right)
\NN
&&
+\frac{2 \nu' e^{\lambda}}{r^2} \xi_{2}
+\frac{4}{3}e^{\lambda -\nu}\varpi^2,
\\
f_4 
&=&
\frac{3e^{\lambda}}{4r^2}\left( k_{2} -\frac{ \nu'}{2} \xi_{2}\right)
-4\pi(p_{0} +\rho_{0})
\Biggl(
\left(\frac{e^{\lambda}}{6} +4\right)\varpi^2 r^2 e^{-\nu}
-\frac{8}{\nu'^2}e^{2\lambda -\nu}\omega\varpi
-\frac{r^{12}}{3}e^{-\nu}\left(\frac{\varpi}{r^8}\right)'\varpi'
\Biggr)
\NN
&&
 -\frac{e^{\lambda-\nu}}{4} \varpi^2 
+\left(\frac{1}{r \nu'}-\frac{1}{2}\right)e^{\lambda -\nu}\omega^2.
\end{eqnarray}

\subsection{Coefficient of $U^{(1)}_{L}$, $\beta_{1}$}
\begin{eqnarray}
 \beta _{1}
&=&
c_{3}
\left[
 32\chi \varpi^2
\left\{
\frac{e^{\lambda/2}}{r^2}
  \left(\frac{e^{\lambda/2}r^2}{\nu'}\right)'
+\left(\frac{8\pi e^{2\lambda}}{(\nu')^2 } \right)( \rho_{0}+p_{0} )
\right\}
\right]
\NN
&-&
 c_{2}
\left[
 \frac{16}{\nu'j^2 r^4}
  (\chi \varpi^2 r^4 e^{-\nu})'
+4 \chi \varpi'\left\{
		 (\omega r^2)' 
		+\frac{2e^{\lambda}}{\nu'}\omega
               \right\}
\right]
\NN
&+&
c_{1}
\Biggl[
 \frac{12 \chi}{jr}
  \left(\frac{\xi_{2}}{jr}\right)'
-18\frac{\xi_{2} \chi'}{j^2 r^2}
-12\frac{\chi k_{2} }{j^2 r^2}
\NN
&&
+\frac{120 W_{3}}{l(l+1)j^2 r^2}
+4\chi \varpi^2 e^{\lambda}\left\{ 1 +4\pi( \rho_{0}+p_{0} )r^2 
 \right\}
+\frac{2}{3r^6}\chi \varpi' (\omega r^8)'
\Biggr]
\NN
&+&
\frac{m^{2}}{l(l+1)}
\left[
 6\frac{ (\chi r^2)' }{j^2 r^4}\xi_{2}
+12\frac{\chi k_{2} }{j^2 r^2}
-\frac{120 W_{3}}{l(l+1)j^2 r^2}
+\frac{4}{3}\chi \varpi'(\omega r^2)'
\right]
\NN
&+&
 2\chi (e^{\nu})'
  \left\{h'_{0} +\frac{1}{j^2}
\left ( \frac{m_{0}}{r e^{\nu}} \right )' \right\}
+\frac{ 8W_{1} +48W_{3} }{l(l+1)j^2 r^2}
+\frac{32\pi}{3}\chi \varpi^2 p'_{0} r^3 e^{\lambda}
-\frac{4}{3}\chi \varpi'(\omega r^2)'.
\NN
\end{eqnarray}

\subsection{Coefficient of $\Phi'^{(1)}_{L}$, $ \beta_2 $}
\begin{eqnarray}
 \beta _{2}
&=&
4 c_{2} \chi 
\left[ 4\pi (\rho_{0} +p_{0} ) 
\left\{
 4 \varpi (\omega r)'
- \varpi '\omega r  
\right \} r^3 e^{-\nu} 
 -  \varpi' (\omega r)' j^2 r^2 
\right ]
\NN
&+&c_{1}
\chi r^2 e^{-\nu}     
\Biggl[  
 - \frac{5\pi}{2}( \rho'_{0}+3p'_{0} ) \varpi^2 r^2
+ \varpi'   
\left \{
 \left( \frac{5}{4 \nu' r } - \frac{5}{24} \right) \omega  
 +\left( \frac{14 }{3}  \omega - \frac{97}{24} \varpi' r \right )
 e^{-\lambda} 
\right \}
\NN
&&
-4\pi( \rho_{0}+p_{0} ) 
\left\{
  \frac{5}{2} \varpi^2 r
   +\frac{44}{3} \varpi \omega r
   -\frac{85}{6} \varpi' \varpi r^2
   -\frac{14}{3} \varpi' \omega r^2
   +\frac{6}{\nu'^2 j^2 }  \varpi ( \omega e^{-\nu})'
\right\}
\Biggr]
\NN
&+&m^{2}
\Biggl[
4\pi(\rho'_{0}+3 p'_{0}) \chi \varpi r^4 e^{-\nu}
\left(
 \frac{1}{4}\varpi - \frac{11}{16} \chi
\right)
\NN
&&
+\chi ' \omega e^{-\nu} \left \{
 \frac{ (r^2 e^{\nu} )'}{ 4 \nu' }  \varpi e^{-\nu}
 +   \frac{3}{8 \nu'}  \chi  r
 - \frac{19}{48} \chi  r^2 
\right \}
-\chi j^2 r^2 
\left \{
  \frac{1}{4} \varpi'^2 r 
 -\frac{2}{3}  \chi'  \omega 
 -\frac{1}{48} \chi' \varpi'  r
\right \}
\NN
&&
+ 4\pi( \rho_{0}+p_{0} ) \chi r^2 e^{-\nu}
\left \{
  (\varpi r)' \varpi r
 - \frac{11}{4} \chi \varpi r 
 - \frac{8}{3}  \chi \omega r
 + \frac{1}{j ^2 \nu'^2}  \chi  (\omega e^{-\nu})'
 - \frac{1}{12} \chi  \varpi' r^2
  + \frac{2}{3} \chi' \omega  r^2
\right \}
\Biggr]
\NN
&+&
\left( \chi - \varpi + \frac{\nu'}{2}  \chi'r^2 e^{-\lambda} 
\right )
  \left( h_0' + \left(\frac{  e^{\lambda} m_0}{r}\right)' \right)
+ \frac{2\pi}{3} ( \rho'_{0} +3 p'_{0} ) ( 4\chi -\varpi)
   \varpi^2 r^4 e^{- \nu}
\NN
&&
-(\lambda'+ \nu' ) j^2
\left[
 \frac{1}{9}((\varpi r)^3)'
   +\frac{4}{3}\left(\frac{\varpi}{r}\right)' \chi \varpi r^4
   +\frac{2}{3}\left(\frac{\varpi}{r^4}\right)'\chi \omega r^7
   +\frac{\nu'}{6} \varpi^2  \chi'r^4
  +\frac{\nu' r}{2j^2} \chi ' m_0
\right ]
\NN
&&
+
\left[
\frac{2}{3} \chi 
-\frac{1}{2}  \varpi 
-\frac{1}{\nu' r}\varpi 
\right ] \varpi' \omega r^2 e^{- \nu}
 +\frac{1}{6}  
( \varpi'^2 r - 8\chi' \omega +4
 \chi ' \varpi' r ) \varpi  j^2 r^2.
\end{eqnarray}

\subsection{Coefficient of $\Phi^{(1)}_{L}$, $\beta_{3}$}
\begin{eqnarray}
 \beta_{3}
&=&
c_{2}
\Biggl[ 
16\pi r^2e^{-\nu} \chi \varpi (r^2 \omega)' (\rho'_{0} +3p'_{0})
-16\pi(\rho_{0} +p_{0})e^{-\nu}\chi
 \Biggl(
16\pi(\rho_{0} +p_{0})r^4e^{\lambda } \varpi(\varpi +\Omega)
\NN
&&
 +2r^2e^{\lambda}\Omega\varpi
  \left(1 +\frac{8}{r\nu'} -\frac{32\pi}{\nu'^2}e^{\lambda}p_{0}\right)
 -6 r \omega( r^2 \varpi )'
 -\frac{2re^{\lambda}}{\nu'^2}\varpi'(
   r\nu'\varpi+2\Omega -r\varpi' )
\NN
&&
 +r^3\varpi'(4\varpi+5r\varpi')
\Biggr)
+\frac{4}{r \nu'}(r^2 e^{-\nu})'\chi\omega\varpi'
-4e^{-\nu} r^2 \chi' \varpi'
( 4\pi(\rho_{0} +p_{0})r^2\omega +e^{-\lambda}(r\omega)')
\Biggr]
\NN
&+&c_{1} \Biggl[
\frac{16\pi e^{\lambda -\nu}}{\nu'^2}(\rho_{0} +p_{0})\Omega
 \Biggl\{
 112\pi r^2e^{\lambda} \varpi \chi p_{0}
 -4e^{\lambda}\chi\omega
 -2r \nu' (9\varpi +\omega)\chi
 +6 r\chi\varpi'
 +7 r^2 \nu' \chi\varpi'
\NN
&&
 +r^2 \nu' \omega\chi'
\Biggr\} 
 +2 re^{-\nu} \omega\chi' 
\left( 2\varpi+ \left(1 - \frac{2}{r\nu'} \right) \omega \right)
 - r^2 e^{-\nu}\varpi'\chi' 
\left(\varpi- \left(1 +\frac{2}{r\nu'}  \right)\omega \right)
+ g_1
\Biggr]
\NN
&+&\frac{m^2}{l(l+1)}\Biggl[
-4\pi (\rho_{0} +p_{0})e^{\lambda-\nu} \varpi
\biggl( \frac{23}{4}r^2\varpi^2
  +\frac{15}{4}r^2\varpi\omega
  +\frac{4e^{\lambda}}{\nu'^2}\Omega\omega
\biggl)
+\frac{e^{\lambda -\nu}}{2}( 5\varpi+\omega )
\left( 1- \frac{2}{r \nu'} \right)\omega^2
\NN
&&
+\frac{16\pi}{\nu'}(\rho'_{0} +3p'_{0})
r^2 e^{\lambda -\nu} \Omega^2 ( \varpi+\Omega)
-32\pi(\rho_{0} +p_{0})\varpi \chi  e^{\lambda-\nu}
\biggl(
( \Omega +\omega)r^2 
+\frac{\Omega}{\nu'^2 }( 1- e^{\lambda})
\biggr)
\NN
&&
+r^2\Omega\varpi'^2e^{-\nu}
+ 4 \chi e^{-\nu} \left(e^{\lambda} \varpi^2
  -\frac{2}{r\nu'}\omega^2 \right)
-g_1+3g_2 -g_3 
\Biggr]
-m^2\left(1-\frac{2}{r \nu'}\right)e^{\lambda-\nu}\Omega \omega^2
\NN
& + &
\Biggl[
l(l+1)(\Omega - \chi )\Biggl\{
\left(1 -\frac{2}{\nu' r}\right)e^{\lambda-\nu}\omega^2
 -\frac{2e^{2\lambda}}{r^3}m_{0}
\Biggr\}
+\frac{8\pi}{3}(\rho_{0} +p_{0})r^2e^{\lambda -\nu}\varpi
\{ \varpi^2 -4 ( \varpi-\omega) \chi \}
\NN
&&
+\Omega\Biggl\{
-4\pi e^{\lambda}(\rho_{0} +p_{0})\varpi
\biggl(
\frac{r^2}{12}e^{-\nu}\varpi
+\frac{4}{\nu'^2}e^{\lambda -\nu}\omega
\biggr)
+\frac{4}{r^3}e^{2\lambda} m_{0}
- \nu'\left(\left(\frac{e^{\lambda}}{r}m_{0}\right)' +h'_{0}\right)
\Biggr\}
\NN
&&
+\frac{4}{r^3}e^{2\lambda}
\{ 4\pi ( 8\pi (\rho_{0} +p_{0})r^2\chi +\chi-\Omega )p'_{0} r^3 -\chi \}
  \left( m_{0} + \frac{r^3}{3} j^2 \varpi^2 \right)
\NN
&&
+\frac{1}{r} \{ (4  +\nu' r -16\pi p'_{0}r^3 )\chi +\chi' r\}
\left(\left(\frac{e^{\lambda}}{r}m_{0}\right)' +h'_{0}\right)
-g_2 +g_3
\Biggr],
\end{eqnarray}

where
\begin{eqnarray}
l(l+1)g_1 &=& 
4\pi (\rho'_{0} +3p'_{0})\Biggl\{
\frac{3}{4}(8\varpi +3\Omega)e^{\lambda}\xi_{2}
+\frac{r^2\Omega}{\nu'}e^{\lambda-\nu}(\varpi +\Omega)(4\varpi +3\Omega)
\NN
&&
+\frac{5}{6}e^{-\nu}r^2 \varpi^2 ( 4(r^2\varpi)' -11\Omega r)
\Biggr\}
+\frac{20\pi}{3}(\rho_{0} +p_{0})^2e^{\lambda -\nu}r^4\varpi^2
 (8\varpi +11\Omega)
\NN
&&
+4\pi(\rho_{0} +p_{0}) e^{\lambda}
 \Biggl\{
 -\frac{3}{2r} (5\varpi-3\omega) \xi_{2}
 +\frac{5}{12}r^2 j^2 (8\varpi -11\Omega) 
    (12 \varpi^2 -(r\varpi')^2 )
\NN
&&
 +\frac{44}{3}r^2 j^2 \varpi \varpi'(r^2 \varpi )'
 -26r^3j^2\Omega \varpi\varpi'
+ \frac{17}{3}e^{-\nu} r^2 \varpi^3
 - \frac{1123 }{24}e^{-\nu} r^2 \Omega \varpi^2 
\NN
&&
 +\frac{\Omega e^{-\nu}}{\nu'^2}
(  -22 e^{\lambda } \varpi \omega
 +2r\nu' ( 20 \varpi^2 +12 \varpi \omega+3\omega^2 )
+  8 r  \varpi \varpi'
+  (32\varpi+13 \omega) r^2  \nu'\varpi')
 \Biggr\}
\NN
&&
+(5\varpi -3\omega)\Biggl\{
 \left(\frac{45}{16 r^2}  -\frac{6}{\nu' r^3} \right)e^{\lambda}k_{2}
-\frac{3}{2 r^3}
 \left(1 -3e^{\lambda} -\frac{\nu' re^{\lambda}}{16} \right)\xi_{2}
+\frac{e^{\lambda -\nu}}{16}\varpi^2
\NN
&&
 + \left(1 - \frac{2}{ r \nu'}\right)\frac{e^{\lambda -\nu}}{8}\omega^2 
 -\frac{2e^{-\nu}}{r \nu'}\omega^2
+ (3e^{\lambda }-1)
 \frac{\varpi^2 e^{ -\nu}}{r \nu ' }
 - \frac{r}{4 \nu ' }e^{ -\nu} (\varpi')^2
\Biggl\} 
-\frac{7}{\nu'}e^{-\nu}\Omega\omega\varpi'
\NN
&&
  + \frac{re^{-\nu}}{6}( 23 \varpi  -9\omega) \omega \varpi'
  + \frac{r^2e^{-\nu}}{96} ( 437 \varpi -403 \omega )\varpi'^2
+\frac{5}{12}r^2 j^2\varpi'^2 ( 11\Omega -8(r\varpi)'),
\NN
\\
g_2 &=&
l(l+1)e^{\lambda}\Omega\Biggl\{
\frac{4\pi}{3}(\rho_{0} +p_{0})r^2 e^{-\nu}\varpi^2
-\frac{1}{2 r^2}\left( k_{2}  -\frac{\nu'}{2}\xi_{2} \right)
+\frac{e^{-\nu}}{6}\varpi^2
+\frac{r^2 j^2}{12}\varpi'^2
\Biggr\}
\NN
&&
+4\pi(\rho'_{0} +3p'_{0}) e^{\lambda} 
\Biggl\{ \left( 2  \chi +\frac{\Omega  }{2} \right) \xi_{2} 
+\frac{j^2}{3}\varpi
( r^3 \varpi ( \Omega- 4\chi) -2\chi ( r^4 \omega)' )
\NN
&&
-\frac{2}{\nu'}r^2e^{-\nu}\Omega^2 (\varpi + \Omega) 
\biggr)
\Biggr\}
 -\frac{32\pi^2}{3}(\rho_{0} +p_{0})^2 r^4
 e^{\lambda -\nu}\varpi (  \Omega\varpi -4 \chi ( \varpi + 2\Omega) )
\NN
&&
+4\pi(\rho_{0} +p_{0})\Biggl\{
e^{\lambda} \Omega \biggl(
\frac{\xi_{2}}{r} +\frac{ r^2 j^2}{6} (12\varpi^2
  +8 r \varpi'\varpi  - ( r \varpi' )^2 )
\biggr)
\NN
&&
 -\frac{4}{r} e^{\lambda}\chi \xi_{2}
   -\frac{2}{3}r^2 e^{ -\nu}\biggl( 
    12(\varpi+2 \omega ) \varpi \chi 
    +4 r (\varpi+3\omega ) \chi \varpi' 
    -(5\varpi+2\omega ) r^2 \varpi' \chi'
\biggr)
\Biggr\}
\NN
&&
-\frac{1}{r^3}\biggl( 
 r \nu' e^{\lambda} \left(  \frac{\varpi}{2}
   +\frac{29}{16} \Omega -2 \chi \right)
-(1 -3e^{\lambda}) ( \Omega -4 \chi )
\biggr)\xi_{2}
\NN
&&
+\frac{e^{\lambda}}{r^2}
 \biggl( \varpi -\frac{19}{8}\Omega +8\chi 
+\frac{4}{r \nu'}( \Omega - 4\chi )
\biggr)k_{2}
-\frac{r^2}{6} \varpi'^2 e^{-\nu} \biggl(
e^{-\lambda}\Omega +\frac{47}{8}\Omega 
+ \varpi  -\frac{\Omega}{r \nu'}
\biggr)
\NN
&&
-e^{\lambda-\nu}\varpi^2 \biggl(
   \frac{\varpi}{3}+ \frac{29}{24}  \Omega
   +\frac{2}{3\nu' r}(3-e^{-\lambda})( \Omega-4\chi  )
\biggr)
\NN
&&
+\frac{2r}{3\nu' }e^{-\nu}\chi'\Biggl\{
   e^{-\lambda}\nu'\varpi' (r\varpi+( r^2 \omega )')
  -\varpi\varpi'
  +  2 \nu' \varpi\omega
   + r\nu' \varpi'(3 \varpi -\omega)    \Biggr\},
\\
g_3 &=& 
\Omega e^{-\nu}\Biggl\{
\frac{8\pi re^{\lambda}}{\nu'}(\rho_{0} +p_{0})
\biggl(
2(\Omega +\varpi )^2
+r\varpi'(3\Omega+\varpi)
\biggr)
\NN
&&
-\frac{e^{\lambda}}{4}\left( 1- \frac{2}{r \nu' }\right)\omega^2
+\frac{4}{\nu' r} \omega^2
-\left(r +\frac{2}{\nu'}\right)\omega\varpi'
\Biggr\}.
\end{eqnarray}

\subsection{${\cal G}[H^{(1)}_{0,\pm}, K^{(1)}_{\pm}]$}
\begin{eqnarray}
{\cal G}[H^{(1)}_{0,\pm}, K^{(1)}_{\pm}]
&=&
A_{1}\partial_{T}K^{(1)}_{\pm}
+B_{1}\partial_{T}H^{(1)}_{\pm}
+im A_{2}K^{(1)}_{\pm}
+im B_{2}H^{(1)}_{\pm},
\end{eqnarray}
where
\begin{eqnarray}
A_{1}
&=&
Q_{\pm}\Biggl[
\frac{2e^{\lambda}}{r^2}\omega
+64\pi (\rho_{0} +p_{0})\left(\frac{e^{-\nu}}{j^2 r \nu'}\right)^2\varpi
\Biggr]
\NN
&&
+S_{\pm}\Biggl[
16\pi e^{\lambda}(\rho_{0} +p_{0})
\left\{\chi -\frac{2e^{\lambda}}{r^2\nu'^2}(\varpi +\chi)\right\}
-\frac{2e^{\lambda}}{r^2}\left(1 -\frac{2}{r\nu'}\right)\omega
\Biggr]
+T_{\pm}\Biggl[
-\frac{e^{\lambda -\nu}}{\nu'}\left(\frac{e^{\nu}}{r^2}\right)'\omega
\Biggr],
\NN
\\
B_{1}
&=&
Q_{\pm}B_{3}
+S_{\pm}\Biggl[
\frac{16\pi e^{\lambda}}{r^2 \nu'^2}(\rho_{0} +p_{0})
\Biggl\{
2\chi(e^{\lambda} +r\nu') +2e^{\lambda}\varpi
-r^2\nu'\varpi'
-(2\varpi +\omega)\left(1-\frac{3}{4}r\nu'\right)r\nu'
\Biggr\}
\NN
&&
\hspace*{1cm}
-(2\varpi +\omega)
 \left(\frac{8\pi e^{-\nu}\rho'_{0}}{j^2 \nu'}\right)
-\frac{4}{r^3 \nu'}\omega
\Biggr]
+T_{\pm}\Biggl[
-\frac{2e^{\lambda}}{r^3 \nu'}\left(
1 -\frac{r \nu'}{2}
\right)\omega
\Biggr],
\\
A_{2}
&=&
Q_{\pm}\Biggl[
+2\chi e^{\lambda}
\left(8\pi \Omega (\rho_{0} +p_{0}) -\frac{\omega}{r^2}\right)\Biggr]
+S_{\pm}\Biggl[
-2\chi e^{\lambda}
\left\{8\pi \Omega (\rho_{0} +p_{0}) 
 -\left(1-\frac{2}{\nu'r}\right)\omega\right\}
\Biggr]
\NN
&&
+T_{\pm}\Biggl[
+\chi e^{\lambda}
\left(\frac{1}{r^2} -\frac{2}{\nu'r^3}\right)\omega\Biggr],
\\
B_{2}
&=&
Q_{\pm}\Biggl[
-\chi B_{3}
+8\pi\chi e^{\lambda}(\rho_{0} +p_{0})\varpi\Biggr]
+S_{\pm}\Biggl[
\chi B_{3}
-2\chi e^{\lambda}
\left\{\frac{e^{-\nu}\omega}{\nu'}
 \left(\frac{e^{\nu}}{r^2}\right)'
 +4\pi(\rho_{0} +p_{0})\varpi\right\}\Biggr]
\NN
&&
+T_{\pm}\Biggl[
\frac{e^{\lambda -\nu}}{\nu'}\left(\frac{e^{\nu}}{r^2}\right)'\chi\omega
+4\pi(\rho_{0} +p_{0})
\left\{
\frac{(r^4 \chi^2 e^{-\nu})'}{j^2r^4\nu'}
-8\left(\frac{\chi e^{-\nu}}{r\nu'}\right)^2
\right\}\Biggr],
\\
B_{3}
&=&
(2\varpi +\omega)\left\{
\frac{8\pi e^{-\nu}\rho'_{0}}{j^2 \nu'} 
+4\pi e^{\lambda}(\rho_{0} +p_{0})\left(\frac{4}{r \nu'} -3\right)\right\}
+\left(
\frac{4}{r^3 \nu'}(1-e^{\lambda}) +\frac{2 e^{\lambda}}{r^2}
\right)\omega
\NN
&&
+\frac{16\pi e^{\lambda}}{\nu'}(\rho_{0} +p_{0})
 \left(\varpi' -\frac{4 e^{-\nu} \varpi}{j^2 r^2 \nu'}\right).
\end{eqnarray}

\end{document}